\documentclass[prd,aps,preprint,tightenlines,nofootinbib,superscriptaddress]{revtex4}

\usepackage{epsfig}

\usepackage{amsmath}

\input{epsf}

\begin{document}


\vspace*{2cm}

\title{Transverse single spin asymmetry in hadronic $\eta_{c,b}$ production }

\author{ Andreas Sch\"afer, Jian Zhou
 \\[0.3cm]
{\normalsize\it Institute for Theoretical Physics, Regensburg University,} \\
{\normalsize\it Universit\"atsstra{\ss}e 31, D-93053 Regensburg, Germany} \\[0.15cm]
}
\begin{abstract}

We study  the transverse single spin asymmetry in $\eta_{c,b}$
production in polarized hadron collisions, employing the collinear
twist-3 approach in combination with the color singlet model. Our
main focus lies on the contribution from the twist-3
Efremov-Teryaev-Qiu-Sterman function. By extrapolating the derived
spin dependent cross section to the small transverse momentum
region, consistency between the collinear twist-3 approach and the
transverse momentum dependent factorization approach is confirmed.
As a byproduct of this work, we identify a term contributing
to the scale evolution of tri-gluon correlations in the
flavor-singlet case which was previously missed.
\end{abstract}

\maketitle

\section{Introduction}
Attempts to understand the large transverse single spin asymmetries (SSAs) observed in the various high energy scattering
processe presently form an active field of research.
The experimental observations are a great theoretical challenge as
the collinear leading twist contribution to the asymmetries is proven to be
proportional to the quark mass~\cite{Kane:1978nd,Ma:2008gm} and thus very small.
However, during the past few decades, remarkable theoretical progresses was achieved
by following mainly two approaches: one approach is based on transverse momentum dependent(TMD)
factorization~\cite{Sivers:1990fh, Boer:1997nt, Collins:1992kk, Brodsky:2002cx,Collins:2002kn,Metz:2002iz}
and the other on collinear twist-3
factorization~\cite{Efremov:1981sh,Qiu:1991pp,Qiu:1998ia,Eguchi:2006qz,Kouvaris:2006zy,Ji:2006ub,Ji:2006vf,Zhou:2009jm,Koike:2009ge}.
In TMD factorization, a naive time reversal odd TMD distribution, known as Sivers function~\cite{Sivers:1990fh}
describing the correlation between the parton
intrinsic transverse momentum and hadron transverse spin vector is responsible for the asymmetries.
In the collinear twist-3 approach
the SSAs arise from twist-3 quark gluon correlators
so-called Efremov-Teryaev-Qiu-Sterman functions (ETQS) $T_F(x,x)$~\cite{Efremov:1981sh,Qiu:1991pp}.
These two framework have their own kinematic regions of validity and were shown to produce the same results in
the overlap of these kinematic regions~\cite{Ji:2006ub,Ji:2006vf,Zhou:2009jm,Bacchetta:2008xw}.

The fundamental property of both mechanisms is that the imaginary part necessary
for non-vanishing SSAs is
dynamically generated through initial/finial state interactions.
In the case of TMD factorization, the effect of
these initial/finial state interactions is encoded in the process dependent gauge link appearing the matrix
element defining the TMD distributions.
Due to different structures of these gauge links in different processes, naive time reversal odd TMD distributions, like
the Sivers function
possess a very unique, modified universality property, namely,
that the Sivers functions in DIS and Drell-Yan process differ by a minus sign~\cite{Brodsky:2002cx,Collins:2002kn}.

While SSAs in DIS have been measured by HERMES~\cite{Diefenthaler:2007rj} and COMPASS~\cite{Martin:2007au}, a
polarized Drell-Yan measurement is not yet available, such that this prediction could not yet be tested.
Alternatively, one can simply fit the ETQS function related to the $k_T$ moment of the Sivers function~\cite{Boer:2003cm}
using input from HERMES and COMPASS~\cite{Anselmino:2008sga} and compare it with a direct extraction of $T_F$
from the SSA obserevd for pion production
in polarized $pp$ collisions~\cite{Adams:1991rw,Adams:2003fx,Arsene:2008aa,Kouvaris:2006zy}.
Promising early results along these lines, have, unfortunately been caused by a sign error, such that
the obserevd discrepancies are now referred to as
``sign mismatch''  of these two processes~\cite{Kang:2011hk}.
Our recent work~\cite{Metz:2012ui} also indicated that this procedure gives the wrong sign for
SSA in inclusive DIS off a neutron target
(for a complete theoretical treatment of this asymmetry, see Ref.~\cite{Schlegel:2012ve})
if the quark gluon correlation $T_F$ fitted to data for the process $p\uparrow p\rightarrow \pi X$ were used as the input
for the twist-3 quark photon correlation.
This suggests that the SSAs observed in process like $p\uparrow p\rightarrow \pi X$ are not
mainly caused by the Sivers mechanism as described in the twist-3 collinear
approach~\cite{Qiu:1998ia,Kouvaris:2006zy}. A possible explanation is suggested by the observation~\cite{Anselmino:2012rq}
that the Collins effect~\cite{Collins:1992kk, Yuan:2009dw} describing SSA as caused by parton fragmentation
can contribute significantly. This observation has been confirmed by a more recent analysis~\cite{Gamberg:2013kla}.

In order to better understand the Sivers mechanism in hadronic collisions, it is desirable to
investigate SSA for cases of particle production in polarized $pp$
collisions for which the Collins effect is absent. Possible options are the SSA in direct photon or jet production
production $p\uparrow p\rightarrow \gamma(Jet) X$~\cite{Qiu:1991pp,Nogach:2012sh}.
Another option which was initially proposed in Ref.~\cite{Yuan:2008vn}  is the SSA in heavy quarkonium production.
The author of the paper~\cite{Yuan:2008vn}
analyzed SSA in heavy quarkonium production following the general arguments of
non-relativistic QCD(NRQCD)~\cite{Bodwin:1994jh,Brambilla:2010cs}.
In the framework of NRQCD, the heavy quarkonium can be produced at short distance
not only in the color singlet but also the color octet configuration.
The transition from heavy quark and antiquark pair to a quarkonium state is
treated as a non-perturbative process
and encoded in the long distance universal matrix elements
which are characterized according to the velocity expansion of NRQCD.
As demonstrated in Ref.~\cite{Yuan:2008vn}, the SSA in heavy quarkonium production offers
a unique way to investigate its production mechanisms since the SSA crucially depends on the
final state interactions which differ significantly for heavy quarkonium
production in the color singlet and the color octet channel.
Recently, a SSA in $p\uparrow p\rightarrow J/\psi X$ has been measured at PHENIX
and found to be sizeable~\cite{Adare:2010bd}.
According to the analysis of Ref.~\cite{Yuan:2008vn}, this hints at a dominance of a color singlet mechanism
at low transverse momentum and at a non-zero gluon Sivers effect.
Also, in Ref.~\cite{Godbole:2012bx}, the color evaporation model in combination with TMD factorization was used
to study the SSA for $J/\psi$ production in $ep\uparrow$ collisions.

Following this research line, we apply the collinear twist-3 approach to compute the
SSA for  heavy quarkonium production in polarized $pp$ collisions.
In the present work,  the earlier pioneering analysis~\cite{Yuan:2008vn} is extended and refined in the sense
that we compute the transverse momentum dependent behavior of the spin asymmetry, and also take into account
the hard gluon pole contribution.
We only focus on $\eta_{c,b}$ production in this paper, though our formalism can be
easily extended to SSAs for the production of other C-even quarkonia, such as $\chi_{c0}$, $\chi_{b0}$, $\chi_{c2}$ and $\chi_{b2}$.
In leading order, a C-even quarkonium can be produced through two gluon fusion,
which makes it a promising way to access both unpolarized and linearly polarized gluon TMD
distributions~\cite{Brodsky:2012vg,Lansberg:2012kf,Rakotozafindrabe:2013au,Diakonov:2012vb,Boer:2012bt}.
The application of TMD factorization in these processes has been justified by a recent
NLO calculation~\cite{Ma:2012hh}.
Experiments measuring such  $\eta_{c,b}$ asymmetries  could be performed at RHIC and a proposed fixed target experiment
at LHC(AFTER)~\cite{Brodsky:2012vg,Lansberg:2012kf,Rakotozafindrabe:2013au}.
LHC would allow us to study the gluon polarization effect in
a deep saturation regime\cite{Metz:2011wb} due to the very high energy of LHC and relative low mass of C-even charmonium.
In Refs.~\cite{Maltoni:2004hv,Hao:2006nf}, the promising channels to detect $\eta_b$ were discussed.

The analysis~\cite{Bodwin:1994jh} based on NRQCD suggested that for C-even quarkonium production,
the color octet contribution is suppressed
while the color singlet contribution dominates, in particular at low transverse momentum~\cite{Hagler:2000dd},
though complications may arise at large transverse momentum~\cite{Mathews:1998nk,Biswal:2010xk}.
Moreover, C-even bottomonium production in the color octet configuration can certainly
be neglected~\cite{Brodsky:2012vg,Lansberg:2012kf,Maltoni:2004hv},
since it is strongly suppressed in the velocity expansion.
On the other hand, the SSA in C-even heavy quarkonium production receives a contribution not only
from ETQS function but also from the tri-gluon correlation~\cite{Ji:1992eu}.
Nevertheless, we restrict ourselves to color singlet production, i.e. the $T_F$ contribution, and leave
the color octet part and tri-gluon correlation for a future study,
as we are primarily aiming at establishing a formalism that combines collinear twist-3 techniques
and NRQCD factorization to describe SSA in heavy quarkonium production in the present paper.

Our calculation is carried out in the covariant gauge closely following the techniques
outlined in Refs.~\cite{Qiu:1991pp,Qiu:1998ia,Kouvaris:2006zy}
in which the SSAs for pion and direct photon production in hadronic collisions were computed.
Recently, in analogy to the SSAs, the double spin asymmetry $A_{LT}$
in the same processes was also studied within the collinear twist-3 framework~\cite{Liang:2012rb}.
Unlike the SSA in pion or photon production, TMD factorization could be applied in the kinematical region where
the transverse momentum of the heavy quarkonium is much smaller than its mass.
We extrapolate the complete collinear twist-3 result
to the small transverse momentum region and find that it is consistent with the result obtained from TMD factorization.
The key step in establishing this connection is to derive the large transverse momentum gluon Sivers function
and relate it to the ETQS function.
Doing so, we find that the hard gluon pole contribution to the gluon Sivers function was
overlooked in the previous literature~\cite{Yuan:2008vn}.
We will also comment on the scale evolution of tri-gluon correlations for the flavor singlet case in the following section.

The paper is organized as follows: in the section 2, we first briefly review $\eta_{c,b}$ production in unpolarized hadron collisions.
Next we derive the spin dependent differential cross section for $\eta_{c,b}$ production in hadronic collisions,
presenting the result in the limits of high and low transverse momentum.
In particular, we discuss the matching between TMD factorization and the collinear twist-3 approach at small
transverse momentum. The paper is summarized in section 3.

\section{Calculation of the unpolarized and polarized cross sections}
We start by introducing the relevant kinematical variables. For the process under consideration,
\begin{eqnarray}
A(P, \vec S_\perp)+B(P') \rightarrow \eta_{c,b}(l) +X ,
\end{eqnarray}
we define the 4-momenta and polarization vector of the incoming nucleons $A, B$ and outgoing heavy quarkonium $\eta_{c, b}$ as indicated.
The Mandelstam variables are $S=(P+P')^2$, $T=(P'-l)^2$ and $U=(P-l)^2$.
The corresponding Mandelstam variables on the partonic level are given by
$\hat s=(xP+x'P')^2$, $\hat t=(x'P'-l)^2$ and $\hat u=(xP-l)^2$
where $x$, $x'$ are the longitudinal  momentum fractions carried by the partons from nucleon $A$ and $B$ respectively.
The squared invariant mass of the heavy quarkonium is $M^2=l^2$ and $M=2M_Q$ up to small relativistic corrections,
where $M_Q$ is the heavy quark mass.
\begin{figure}[t]
\begin{center}
\includegraphics[width=13cm]{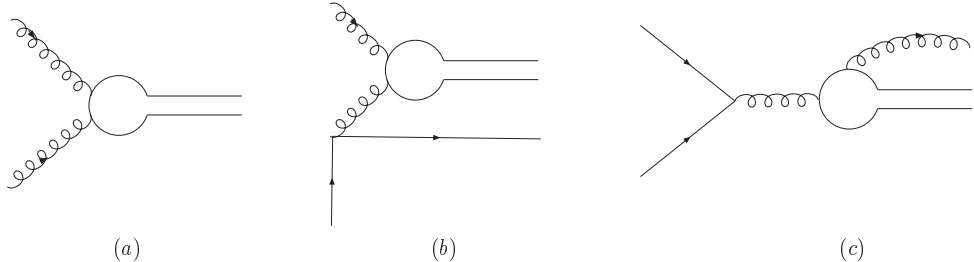}
\caption[] {Diagrams contributing to the unpolarized cross section at order O($\alpha_s^2$) and O($\alpha_s^3$).
  Gluon-gluon scattering  at order O($\alpha_s^3$) is not shown.
Diagrams with permutations of the gluon lines are also not shown.}
\end{center}
\end{figure}
In the unpolarized case, to lowest order $O(\alpha_s^2)$, one only has the gluon fusion process shown in Fig.1(a).
This production mechanism dominates at small transverse momentum where the mean produced quarkonium transverse
momentum is of order $\Lambda_{QCD}$.
At order $O(\alpha_s^3)$, there are much more contributions from the two-by-two scattering processes illustrated in Fig.1(b) and Fig.1(c).
In these processes, the large heavy quarkonium transverse momentum is generated by recoiling against an outgoing gluon or quark.
Note that heavy quarkonium
production at $O(\alpha_s^3)$  from gluon gluon scattering is not shown in the Fig.1.

Within the so-called color singlet model~\cite{Kuhn:1979bb,Guberina:1980dc,Baier:1983va},
the amplitude for the processes discussed has the general form,
\begin{eqnarray}
{\cal M}=\int \frac{d^4 q}{(2\pi)^4} {\rm Tr} [ O(l,q) \phi(l,q)]
\end{eqnarray}
where $2q$ is the relative momentum of the heavy quarks.
Here, $\phi(l,q)$ is the Bethe-Salpeter wave function of the produced bound state.
$O(l,q)$ is the
perturbative part of the diagrams shown in Fig.1 with the heavy quark legs cut off.
Due to the non-relativistic nature of heavy quarkonium,
the relative momentum is assumed to be small with respect to the quark mass $M/2$.
For the production of S-wave states, one can simply neglect the dependence of  $O(l,q)$ on the relative momentum.
This leads to,
\begin{eqnarray}
{\cal M}=\frac{1}{\sqrt{4\pi}} R_0(0) [ O(l,0) {\cal P}_{SS_z}(0)]
\end{eqnarray}

Here, $R_0(0)$ is the value of the radial S-wave function at the origin,
 and ${\cal P}_{SS_z}(0)$ is the spin projection operator. According to the color singlet model, the heavy quark pair produced in the
partonic scattering can evolve into a quarkonium non-perturbatively only if they have the same quantum numbers as
the corresponding quarkonium.
This implies that the spin projector operator associated with $\eta_{c,b}$ production has the form,
\begin{eqnarray}
{\cal P}_{SS_z}(0)=\frac{1}{4 M^{3/2}}
( l \!\!/ +M) \gamma^5 (l \!\!/ +M)
\end{eqnarray}
With these calculational recipes, one derives the unpolarized differential cross section for the gluon fusion channel~\cite{Baier:1983va},
\begin{eqnarray}
\frac{d \sigma}{dy d^2 l_\perp}=\sigma_0
 G_1(z) G_1(z')
\label{Eq:5}
\end{eqnarray}
where $y$ is the rapidity of the produced bound state. $G_1(z), G_1(z')$ are the unpolarized collinear gluon distribution
functions of two incoming hadrons.
The longitudinal momentum fractions are constrained by kinematics: $z=\frac{M}{\sqrt{s}} e^y$, $z'=\frac{M}{\sqrt{s}} e^{-y}$.
In Eq.(\ref{Eq:5}), $\sigma_0$ is given by,
\begin{eqnarray}
\sigma_0=\frac{\pi^2 R^2_0 \alpha_s^2}{3M^3  S}
\end{eqnarray}
For the quark gluon scattering channel, we have~\cite{Baier:1983va},
\begin{eqnarray}
\frac{d \sigma}{dy d^2 l_\perp}=\sigma_0  C_F
\frac{\alpha_s}{2\pi^2}M^2   \sum_{a} \int \frac{dx}{x} \frac{dx'}{x'}
f_1^a(x) G_1(x')
 \frac{ (\hat t -M^2)^2-2\hat s \hat u}{(-\hat t)(\hat t-M^2)^2}
 \delta( \hat s +\hat t+\hat u-M^2)
\end{eqnarray}
where $f_1^a(x)$ is the unpolarized collinear quark distribution. The index $a$ runs over all quark flavors.
For the $q\bar q $ channel, the unpolarized differential cross section reads~\cite{Baier:1983va},
\begin{eqnarray}
\frac{d \sigma}{dy d^2 l_\perp}=\sigma_0 C_F \frac{-8}{3}
\frac{\alpha_s }{2\pi^2}M^2  \sum_{a} \int \frac{dx}{x} \frac{dx'}{x'}
f_1^a(x) \bar f_1^a(x')
 \frac{ (\hat s -M^2)^2-2\hat t \hat u}{-\hat s(\hat s-M^2)^2}
 \delta( \hat s +\hat t+\hat u-M^2)
 \end{eqnarray}
One notices that apart from the color factor, the hard coefficient appearing in the $q\bar q $ channel
can be obtained from those of the quark gluon channel by crossing $\hat t \leftrightarrow \hat s$.
The result for the gluon-gluon scattering channel can also be found in Ref.~\cite{Baier:1983va}.
Next to leading order corrections to C-even hadronic quarkonium production have been calculated
in Refs.\cite{Petrelli:1997ge,Kuhn:1992qw}.
\begin{figure}[t]
\begin{center}
\includegraphics[width=12cm]{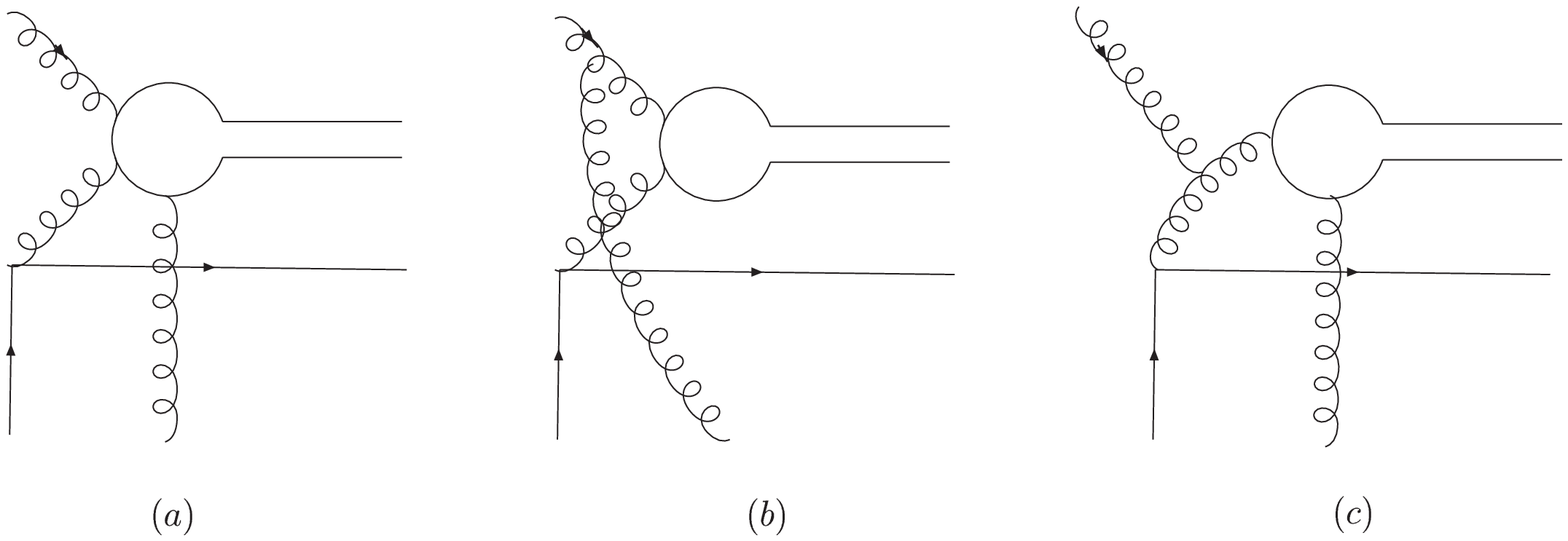}
\caption[] {Sample diagrams giving rise to the SSA in hadronic $\eta_{c,b}$ production from quark gluon channel.}
\end{center}
\end{figure}
Now let's proceed to derive the polarized cross section.
To generate the spin asymmetry, one additional gluon must be exchanged between the active partons
and the remanent part of the nucleon target as shown in Fig.2 and Fig.3.
The hard part, if an additional gluon is attached, can be calculated perturbatively, while
the non-perturbative part describes the relevant three parton correlations.
As stated in the introduction, we only focus on
the contribution in which the quark-gluon correlation in the
transversely polarized nucleon enters. This is precisely the ETQS function which is defined as,
\begin{eqnarray}
&& \int \frac{d\xi^-}{2 \pi} \, \frac{d\zeta^-}{2\pi} \, e^{ix_{1}P^+ \xi^-} e^{i(x-x_1)P^+\zeta^-}
\langle P,S_\perp | \bar{\psi}_\beta(0)  g F_\perp^{+ \mu}(\zeta^-)\psi_\alpha(\xi^{-}) | P,S_\perp \rangle
\nonumber \\
&& \hspace{0.5cm}
=\ \frac{M_N}{2}  T_F(x,x_1) \, \epsilon^{\mu\nu}_\perp S_{\perp \mu} n\!\!\!/_{\alpha \beta}
\end{eqnarray}
where we have suppressed Wilson lines and have indicated the nucleon mass by $M_N$.
We also introduced the lightcone vector $n=(1^+,0^-,\vec 0_\perp)$, whose conjugate
vector is $\bar n=(0^+,1^-,\vec 0_\perp)$.
Note that our definition of the ETQS function differs by a factor $2\pi M_N$ from the conventions used in Refs.~\cite{Ji:2006ub,Ji:2006vf}.
This ETQS function plays an important role for SSA phenomenology. Its scale evolution has been derived in
Refs.~\cite{Kang:2008ey,Zhou:2008mz,Vogelsang:2009pj,Braun:2009mi,Schafer:2012ra,Ma:2012xn,Kang:2012em,Kang:2012ns}.

Similar to the SSA in the Drell-Yan process, the strong
interaction phase factor necessary for having a non-vanishing
spin asymmetry arises from the interference between an
imaginary part of the partonic scattering amplitude with an extra
gluon, as shown in Fig.2 for the quark-gluon scattering channel, and the real
scattering amplitude without a gluon attachment. The
imaginary part is due to the pole of the parton propagator
associated with the integration over the gluon momentum fraction
$x_g $.  This effectively implies that one of the internal parton lines goes on shell.
To isolate the imaginary part of such poles, the distribution identity:
$\frac{1}{x\pm i\epsilon}= {\rm PV} \frac{1}{x}\mp i\pi \delta(x)$ was used.
Depending on which propagator's pole contributes, the amplitude may get contributions from $ x_g=0$ (``soft-pole") and
$x_g\neq 0$ (``hard-pole" )~\cite{Ji:2006ub,Ji:2006vf}. Both types of gluon poles show up in our calculation.

As the reader can find the relevant technical details for our calculation
in the literature~\cite{Qiu:1991pp,Qiu:1998ia,Eguchi:2006qz,Kouvaris:2006zy,Ji:2006ub,Ji:2006vf}
we sketch here only some key steps of such twist-3 calculations.

As mentioned before, we carry out the calculation in the covariant gauge, in which the
leading contribution of the exchanged gluon is the "plus" component $A^+$.
The gluon's momentum is dominated by the $x_gP+k_{g\perp}$, where $x_g$ is the longitudinal momentum fraction
with respect to the polarized proton.
In order to calculate consistently with twist-3 accuracy, one has to expand the hard parts
in the gluon transverse momentum and
keep the terms linear in $k_{g\perp}$.
Then the $k_{g\perp}$ factor can be combined with $A^+$ to yield $\partial^\perp A^+$, which is
an element of the field strength tensor $F^{\partial +}$. After adding the
term proportional to $\partial^+ A^\perp$ of the same tensor~\cite{Eguchi:2006qz,Liang:2008rf},
the soft part can be rewritten in the form of the ETQS function.
\begin{figure}[t]
\begin{center}
\includegraphics[width=14cm]{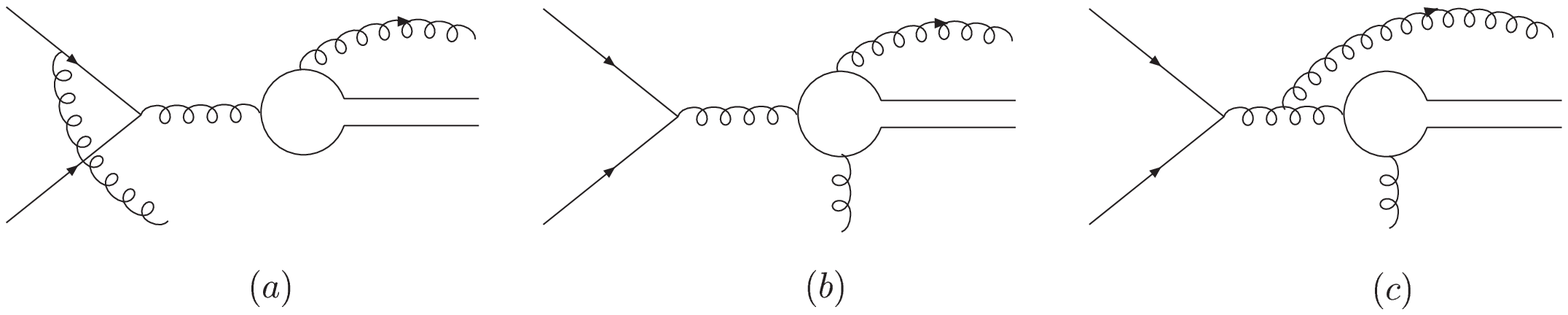}
\caption[] {Sample diagrams giving rise to the SSA in hadronic $\eta_{c,b}$ production from the quark-antiquark channel.}
\end{center}
\end{figure}
Making use of the ingredients described above, the calculation is straightforward.
However, the derived complete results for the spin dependent cross sections is rather length.
To highlight some interesting
features of our results, in the following, we will only present  the expression in
the limit of large transverse momentum ($l_\perp \gg M$) and small transverse
momentum ($\Lambda_{QCD}\ll l_\perp \ll M$) instead of the full expression.
At large transverse momentum, after dropping  all terms suppressed by  powers of $M/l_\perp$,
one ends up with the polarized differential cross section for the $qg$ channel,
\begin{eqnarray}
&& \!\!\!\!\!\!\!\!\!\!\!\!\!\!
\frac{d \sigma}{dy d^2 l_\perp}\approx
\sigma_0 \epsilon^{\mu \nu} S_{\perp \mu }l_{\perp \nu}
\frac{\alpha_s}{\pi}M^2 M_N \sum_{a} \int \frac{dx}{x} \frac{dx'}{x'}\frac{1}{-\hat t} G_1(x')
  \delta( \hat s +\hat t+\hat u)
 \nonumber \\ &\times& \left \{
\left [ 2T_F^a(x,x)-x \left ( \frac{d}{dx} T_F^a(x,x) \right ) \right ]\frac{N_c}{2} \frac{ \hat s^2+ \hat u^2}{-\hat t^3}
\right .\
 \nonumber \\ &&\left .\ -
 T_F^a(x,x) \left ( \frac{N_c}{2} \frac{ 2\hat s^2+\hat t \hat s}{ \hat t^3}-\frac{1}{N_c} \frac{\hat s}{2 \hat u \hat t} \right )
 - T_F^a \left (x,x\frac{\hat s}{\hat s-\hat t} \right )\frac{N_c}{2} \frac{ 2\hat s^3+\hat s^2\hat t-\hat t^3  }{\hat u \hat s  \hat t^2}
 \right \}
\end{eqnarray}
and for the $q\bar q$ channel,
\begin{eqnarray}
&& \!\!\!\!\!\!\!\!\!\!\!\!\!\!
\frac{d \sigma}{dy d^2 l_\perp}\approx
\sigma_0 \left ( \frac{-8}{3} \right )   \epsilon^{\mu \nu} S_{\perp \mu }l_{\perp \nu}
\frac{\alpha_s}{\pi}M^2 M_N \sum_{a} \int \frac{dx}{x} \frac{dx'}{x'}\frac{1}{-\hat t} \bar f_1^a(x')
  \delta( \hat s +\hat t+\hat u)
 \nonumber \\ &\times& \left \{
\left [ 2T_F^a(x,x)-x \left ( \frac{d}{dx} T_F^a(x,x) \right ) \right ]\frac{-1}{2N_c} \frac{ \hat t^2+ \hat u^2}{-\hat s^3}
\right .\
 \nonumber \\ &&\left .\ -
 T_F^a(x,x) \left ( \frac{N_c}{2} \frac{ 2\hat t^2+\hat t \hat s}{ \hat s^3}-\frac{1}{N_c} \frac{\hat t}{2 \hat u \hat s} \right )
 - T_F^a \left (x,x\frac{-\hat t}{\hat s-\hat t} \right )\frac{N_c}{2} \frac{ 2\hat t^3+\hat t^2\hat s-\hat s^3  }{-\hat u \hat t  \hat s^2}
 \right \}
\end{eqnarray}
A few remarks on these analytical results are in order.
First, as usual, the SSA is suppressed by the factor $M_N/l_\perp$ at large transverse momentum.
Second, one notices that the derivative term and non-derivative term from the soft gluon pole
contribution can not be combined into a compact form as it can be done for the polarized cross section
for pion production in hadron collisions~\cite{Kouvaris:2006zy}.
Also, let us note that the hard gluon pole contribution survives even
though the heavy quarkonium mass is neglected, in contrast to the cases of pion and direct photon production for which
 the hard gluon pole contribution is absent.
Finally, it is observed that except for the different color factors,
the soft gluon contribution in the quark gluon channel can be obtained from the quark antiquark channel
by crossing $\hat s\leftrightarrow \hat t$,
while  the hard coefficients associated with the hard gluon pole contributions
for the $qg$ and $q \bar q$ channels differ by a minus sign after crossing
 $\hat s\leftrightarrow \hat t$.

\begin{figure}[t]
\begin{center}
\includegraphics[width=7cm]{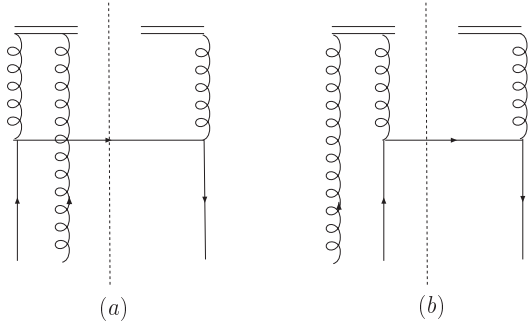}
\caption[] {Diagrams contributing to the gluon Sivers function at large transverse momentum in the flavor singlet case.
Fig.4(a): soft gluon pole contribution; Fig.4(b): hard gluon pole contribution.}
\end{center}
\end{figure}

Now let us discuss our results in the low transverse moment limit.
In this limit, we neglect all terms suppressed by the power of $l_\perp^2/M^2$
while keeping those enhanced by the factor $1/l_\perp^3$.
In order to extrapolate our results to the small transverse momentum region,
one has first to expand the delta function in
$l_\perp$,
\begin{eqnarray}
 \delta( \hat s +\hat t+\hat u-M^2)&=&\delta( \hat s(1-\xi)(1-\xi')-l_\perp^2)
 \nonumber \\
 &=&\frac{1}{\hat s} \left [ \frac{\delta(1-\xi)}{(1-\xi')_+} +\frac{\delta(1-\xi')}{(1-\xi)_+}
 +\delta(1-\xi')\delta(1-\xi) {\rm ln} \frac{M^2}{l_{\perp}^2} \right ]
\end{eqnarray}
where $\xi=z/x$ and $\xi'=z'/x'$. The 'plus' prescriptions is defined in the standard way.
With this expansion, the spin dependent cross section in the kinematical
region $\Lambda_{QCD}\ll l_\perp \ll M$ is given by,
\begin{eqnarray}
&& \!\!\!\!\!\!\!\!\!\!\!\!\!\!\!\!\!\!\!\!\!
\frac{d \sigma}{dy d^2 l_\perp} \approx
\sigma_0  \frac{N_c}{2} \frac{\alpha_s}{\pi}   \frac{M_N}{l_\perp^4}
 \epsilon^{\mu \nu} S_{\perp \mu }l_{\perp \nu} G_1(z')  \int  \frac{dx}{x}
  \nonumber \\ &\times& \!\!\!
\left \{ \left ( x \frac{\partial}{\partial x} T_F(x,x) \right ) \frac{\xi-1}{\xi}[1+(1-\xi)^2]
 -T_F(x,x) 2(1-\xi)^2 -  T_F(x,x-z) \frac{2-\xi}{\xi}  \right \}
   \nonumber \\ &=& \!\!\!
   \sigma_0  \frac{N_c}{2} \frac{\alpha_s}{\pi}   \frac{M_N}{l_\perp^4}
 \epsilon^{\mu \nu} S_{\perp \mu }l_{\perp \nu} G_1(z')  \int  \frac{dx}{x}
\left \{ T_F (x,x) \frac { 1+(1-\xi)^2}{\xi} -T_F(x,x-z)  \frac{2-\xi}{\xi}  \right \}
\end{eqnarray}
In the second step of the derivation of above formula, we carried out the integration over $x$ by parts. It is interesting to note that
the hard coefficient associated with the soft gluon matrix element is the well known splitting kernel ${\cal P}_{gq}$.
Here, it is worthwhile to point out that no leading power contribution at small $l_\perp$
comes from the quark-antiquark channel.

On the other hand, when $l_\perp\ll M$, transverse momentum dependent factorization can be applied.  In the TMD
factorization approach,  the SSA in $\eta_{c,b}$ production is generated by the gluon Sivers function $G_{1T,DY}^\perp$
through the gluon-gluon fusion channel,
\begin{eqnarray}
\frac{d \sigma}{dy d^2 l_\perp}=
\sigma_0 \frac{\epsilon^{\mu \nu} S_{\perp \mu }l_{\perp \nu}}{M_N}
\int d^2k_\perp d^2k_\perp' \frac{k_\perp \cdot l_\perp}{l_\perp^2} \delta^2(l_\perp-k_\perp-k_\perp') G_1(z',k_\perp')G_{1T,DY}^{\perp } (z,k_\perp)
\label{TMD}
\end{eqnarray}
where $G_{1T,DY}^{\perp } (z,k_\perp)$ denotes the gluon Sivers function.  The subscript "DY" indicates that the gluon Sivers
function contains a past-pointing gauge link built up through initial state interactions, similar to that in the Drell-Yan process.

When $k_\perp$ is of the order of $\Lambda_{QCD}$, the gluon Sivers function
is an entirely non-perturbative object. However, in the kinematic region $\Lambda_{QCD} \ll k_\perp \ll M$,
TMD factorization still holds and at the same time the function $G_{1T,DY}^{\perp } (z,k_\perp)$  can be calculated in
terms of twist-three parton correlation function within perturbative QCD.
It can receive contributions from both the ETQS function and tri-gluon correlations.
In the current case, we are interested only in the former one.
Our perturbative calculation follows a similar procedure as in~\cite{Ji:2006ub,Ji:2006vf} resulting in
\begin{eqnarray}
G_{1T,DY}^\perp (z,l_\perp)=
 \frac{N_c}{2} \frac{\alpha_s}{\pi}   \frac{M_N^2}{l_\perp^4}  \int  \frac{dx}{x}
\left \{ _F (x,x) \frac { 1+(1-\xi)^2}{\xi} -T_F(x,x-z)  \frac{2-\xi}{\xi}  \right \}
\label{Gsivers}
\end{eqnarray}
where the soft gluon pole contribution is generated by the diagram shown in the Fig.4(a),
and the hard gluon pole contribution arises from Fig.4(b).
This result was first derived in Ref.~\cite{Yuan:2008vn}, though the second term was missed there.
At this point, we would like to mention that
the hard gluon pole term also contributes to the scale evolution of the tri-gluon correlation for the flavor singlet case,
which was overlooked in the literature previously~\cite{Kang:2008ey,Braun:2009mi}
\footnote{
This term is present in~\cite{Braun:2009mi} in the evolution equation for
arbitrary gluon momenta, but missing in their answer for the gluon
pole limit.}.
In addition, due to the existence of diagram Fig.4(b),
the chiral partner of the ETQS function, $\tilde T_F(x,x_1)$ (notation used in Ref.~\cite{Zhou:2009jm})
also contributes to the scale evolution of tri-gluon correlations~\cite{Manashov}.

By inserting Eq.\ref{Gsivers} into Eq.\ref{TMD} and making the approximation $\delta^2(l_\perp-k_\perp-k_\perp')\approx \delta^2(l_\perp-k_\perp)$,
we reproduce the result obtained in the collinear twist-3 approach.
Therefore, for the observable under  consideration,
we have obtained a unified picture in the kinematical region where TMD factorization and the collinear twist-3 approach both apply,
as was found in many other cases~\cite{Ji:2006ub,Ji:2006vf,Zhou:2009jm}.

To end, let us briefly discuss a possible future numerical study.
In order to estimate the size of SSAs for $\eta_{c,b}$ production,
we must determine the input for the ETQS functions
including the diagonal and off-diagonal contributions, as the polarized cross
section depends on the both, soft and hard gluon pole contributions.
Unfortunately, the off-diagonal contributions to $T_F$  needed for this SSA observable are not
as well-determined as the diagonal pieces which can be related to moments of the quark Sivers function.
In Ref.~\cite{Kang:2008ey}, a Gaussian form was assumed  for $T_F(x,x_{1})$  with a maximum at $x=x_{1}$.
This study was done in the context of the evolution of $T_F(x,x)$.  In Ref.~\cite{Braun:2011aw},
an analysis of higher-twist functions was conducted using light-cone wave functions that
include $qqqg$ Fock states.
In contrast to \cite{Kang:2008ey}, this study found that $T_F(x,x_{1})$  reaches its maximum
when $x\neq x_{1}$ and some of its lowest values when $x=x_{1}$.
These completely different behaviours clearly demonstrate our lack of knowledge
for $T_F(x,x_{1})$.
Nevertheless, it would be useful to determine the potential impact of $T_F(x,x_1)$ on
the size of SSAs in $\eta_{c,b}$ hadronic production.

\section{Summary}
In summary, we have calculated the transverse single spin asymmetry in hadronic $\eta_{c,b}$
production by employing the collinear twist-3 approach in combination
with the color singlet model.  We discussed the behavior of the spin asymmetry
at high and low transverse momentum. In particular, at low transverse momentum,
a match between the TMD factorization approach and the collinear twist-3 formalism has been found
after re-examining the derivation of the gluon Sivers function at large transverse momentum.
As a byproduct of this work, we identified a term, which was missed so far, that contributes also to the scale evolution
of the tri-gluon correlation for the flavor singlet case.
In addition, we have briefly outlined a plan for a future numerical studies.
Let us mention that it would be feasible to measure this asymmetry at RHIC and a proposed fixed target experiment
at LHC(AFTER)~\cite{Brodsky:2012vg,Lansberg:2012kf,Rakotozafindrabe:2013au}.

We emphasize that the transverse spin physics and the heavy quarkonium production physics could mutually benefit from studying this observable as explained in the following.
 First of all, due to the sign mismatch issue~\cite{Kang:2011hk}, a doubt on the validity of the collinear twist-3 approach in hadron collisions has arisen.
 SSA in heavy quarkonium production provides a clean way to test this approach, as compared to that in pion production where the Collins mechanism
 could dominate the asymmetry~\cite{Anselmino:2012rq}.  Furthermore, one notices that the spin asymmetry is independent of the radial wave function at the origin $R_0$
 and only sensitive to the heavy quarkonium production mechanism. Therefore, measuring this observable would provide us an unique chance to pin down the ratio
 between the values of the color singlet long distance matrix element and the color octet matrix element.

There are a number of directions in which our work could be extended. First, the color octet contribution to SSA should be taken into account as it plays a role at
large transverse momentum.  Second, it is natural and straightforward to study the SSAs for other C-even heavy quarkonium
production in the framework outlined in this paper.  It is also possible to study SSAs for hadronic $J/\psi, \Upsilon$ production.
To do so, we have to compute the SSAs generated from the tri-gluon correlation since it is the only contribution appears at the non-trivial leading order in $J/\psi, \Upsilon$
production. Finally, one can calculate the SSAs for photonic/electronic  $J/\psi, \Upsilon$ production using the collinear twist-3 approach in combination with the NRQCD
framework. These would be the relevant observables at the future EIC~\cite{Anselmino:2011ay,Boer:2011fh}.

\

\noindent {\bf Acknowledgments:} This work has been supported by
BMBF under Grant No. OR 06RY9191 and by the NSF under Grant No. PHY-1205942. We thank Alexander Manashov for valuable
discussions on the scale evolution of tri-gluon correlations.

\end {document}